\PassOptionsToPackage{unicode}{hyperref}
\PassOptionsToPackage{hyphens}{url}
\PassOptionsToPackage{dvipsnames,svgnames,x11names}{xcolor}
\documentclass[letterpaper,
12pt]{article}

\usepackage{amsmath,amssymb}
\usepackage{iftex}
\ifPDFTeX
\usepackage[T1]{fontenc}
\usepackage[utf8]{inputenc}
\usepackage{textcomp} % provide euro and other symbols
\else % if luatex or xetex
\usepackage{unicode-math}
\defaultfontfeatures{Scale=MatchLowercase}
\defaultfontfeatures[\rmfamily]{Ligatures=TeX,Scale=1}
\fi
\usepackage{lmodern}
\ifPDFTeX\else  
\fi
\IfFileExists{upquote.sty}{\usepackage{upquote}}{}
\IfFileExists{microtype.sty}{% use microtype if available
	\usepackage[]{microtype}
	\UseMicrotypeSet[protrusion]{basicmath} % disable protrusion for tt fonts
}{}
\makeatletter
\@ifundefined{KOMAClassName}{% if non-KOMA class
	\IfFileExists{parskip.sty}{%
		\usepackage{parskip}
	}{% else
		\setlength{\parindent}{0pt}
		\setlength{\parskip}{6pt plus 2pt minus 1pt}}
}{% if KOMA class
	\KOMAoptions{parskip=half}}
\makeatother
\usepackage{xcolor}
\makeatletter
\ifx\paragraph\undefined\else
\let\oldparagraph\paragraph
\renewcommand{\paragraph}{
	\@ifstar
	\xxxParagraphStar
	\xxxParagraphNoStar
}
\newcommand{\xxxParagraphStar}[1]{\oldparagraph*{#1}\mbox{}}
\newcommand{\xxxParagraphNoStar}[1]{\oldparagraph{#1}\mbox{}}
\fi
\ifx\subparagraph\undefined\else
\let\oldsubparagraph\subparagraph
\renewcommand{\subparagraph}{
	\@ifstar
	\xxxSubParagraphStar
	\xxxSubParagraphNoStar
}
\newcommand{\xxxSubParagraphStar}[1]{\oldsubparagraph*{#1}\mbox{}}
\newcommand{\xxxSubParagraphNoStar}[1]{\oldsubparagraph{#1}\mbox{}}
\fi
\makeatother

\usepackage{longtable,booktabs,array}
\usepackage{calc} % for calculating minipage widths
\usepackage{etoolbox}
\makeatletter
\patchcmd\longtable{\par}{\if@noskipsec\mbox{}\fi\par}{}{}
\makeatother
\IfFileExists{footnotehyper.sty}{\usepackage{footnotehyper}}{\usepackage{footnote}}
\makesavenoteenv{longtable}
\usepackage{graphicx}
\makeatletter
\def\maxwidth{\ifdim\Gin@nat@width>\linewidth\linewidth\else\Gin@nat@width\fi}
\def\maxheight{\ifdim\Gin@nat@height>\textheight\textheight\else\Gin@nat@height\fi}
\makeatother
\setkeys{Gin}{width=\maxwidth,height=\maxheight,keepaspectratio}
\makeatletter
\def\fps@figure{htbp}
\makeatother

\makeatletter
\@ifpackageloaded{caption}{}{\usepackage{caption}}
\AtBeginDocument{%
	\ifdefined\contentsname
	\renewcommand*\contentsname{Table of contents}
	\else
	\newcommand\contentsname{Table of contents}
	\fi
	\ifdefined\listfigurename
	\renewcommand*\listfigurename{List of Figures}
	\else
	\newcommand\listfigurename{List of Figures}
	\fi
	\ifdefined\listtablename
	\renewcommand*\listtablename{List of Tables}
	\else
	\newcommand\listtablename{List of Tables}
	\fi
	\ifdefined\figurename
	\renewcommand*\figurename{Figure}
	\else
	\newcommand\figurename{Figure}
	\fi
	\ifdefined\tablename
	\renewcommand*\tablename{Table}
	\else
	\newcommand\tablename{Table}
	\fi
}
\@ifpackageloaded{float}{}{\usepackage{float}}
\floatstyle{ruled}
\@ifundefined{c@chapter}{\newfloat{codelisting}{h}{lop}}{\newfloat{codelisting}{h}{lop}[chapter]}
\floatname{codelisting}{Listing}

\makeatother
\makeatletter
\@ifpackageloaded{caption}{}{\usepackage{caption}}
\@ifpackageloaded{subcaption}{}{\usepackage{subcaption}}
\makeatother

\ifLuaTeX
\usepackage{selnolig}  % disable illegal ligatures
\fi
\usepackage[]{natbib}
\usepackage{bookmark}

\IfFileExists{xurl.sty}{\usepackage{xurl}}{} % add URL line breaks if available
\hypersetup{
	pdftitle={Title},
	pdfauthor={Author 1; Author 2},
	pdfkeywords={3 to 6 keywords, that do not appear in the title},
	colorlinks=true,
	linkcolor={blue},
	filecolor={Maroon},
	citecolor={Blue},
	urlcolor={Blue},
	pdfcreator={LaTeX via pandoc}}

\RequirePackage{graphicx}
\RequirePackage{booktabs}
\RequirePackage{threeparttable}
\RequirePackage{adjustbox}
\usepackage{pifont}                        % fr checkand cross symbols
\usepackage{arydshln}                      % dashed line in table
\usepackage{amsthm}
\newtheorem*{remark}{Remark}

\newcommand{\anon}{1}

\begin{document}

	\def\spacingset#1{\renewcommand{\baselinestretch}%
		{#1}\small\normalsize} \spacingset{1}

	%%%%%%%%%%%%%%%%%%%%%%%%%%%%%%%%%%%%%%%%%%%%%%%%%%%%%%%%%%%%%%%%%%%%%%%%%%%%%%

	\if1\anon
	{
		\title{\bf Counting on count regression: a reexamination of routinely-cited Negative Binomial specifications}
		\author{Ettore Settanni\thanks{
				Corresponding author. e-mail: \texttt{e.settanni@eng.cam.ac.uk} }\hspace{.2cm}\\
			Department of Engineering, University of Cambridge\\
			and \\
			Reinout Heijungs \\
			School of Business and Economics, Vrije Universiteit Amsterdam \\
			and \\
			Jagjit Singh Srai \\
			Department of Engineering, University of Cambridge
		}
		\maketitle
	} \fi
	
	\if0\anon
	{
		\bigskip
		\bigskip
		\bigskip
		\begin{center}
			{\LARGE\bf  Counting on count regression: a reexamination of routinely-cited Negative Binomial specifications}
		\end{center}
		\medskip
	} \fi
	
	\bigskip
	\begin{abstract}
		Negative Binomial regression is a staple in empirical management research, especially for the analysis of supply chain disruption risks. Its computational structure is often taken for granted: most applications omit the scoring and information equations and defer to a handful of references for details. But what if the evidence provided by those trusted sources disagrees? We reexamine results from a selection of routinely-cited work on Negative Binomial regression, especially with regard to scoring and information equations in the so-called dispersion parameter. For such parameter, we find limitations affecting each stage of the maximum likelihood estimation process, and conclude that there is no reliable expression for the corresponding element of Fisher Information Matrix. For practical relevance, we also look under the hood of an open-source software implementation in \texttt{R}, and show that the notation adopted has some advantages over its published counterparts. Our proposed remediation is simple: to elevate computations that are rarely made explicit. We illustrate our findings in \texttt{R} with the aid of a simplified numerical example that, while obfuscated due to sensitivity, is underpinned by real-world data on clinical trials supply disruptions.
	\end{abstract}
	
	\noindent%
	\vfill

	\newpage
	\spacingset{1.8} % DON'T change the spacing!

	\section{Introduction}\label{sec-intro}	
	There is more to counting than the proverbial means of putting someone to sleep. Patents, customer orders, but also product recalls and supply disruptions are examples of count-based dependent variables measuring outcomes of interest for empirical management research. Well-known specifications for count regression models include Poisson and Negative Binomial, which are named after the probability distributions from which the count-based dependent variables are assumedly drawn. Both are staple techniques in empirical management research, with extensive literature providing overviews \citep[e.g.,][]{GardnerWilliam1995RAoC, Hilbe2014, MartinPeter2021Rmfc} and guidance on model selection \citep{Count_OrgResMet, Ronkko22}. 
	
	Although the Poisson specification is robust to some level of model misspecification  \citetext{e.g., \citealp{Blackburn02072020}, \citealp[][Ch.18]{Wooldridge}}, the Negative Binomial is more prevalent in practice as seldom are empirical data equidispersed. Appendix \ref{secA1} summarizes example implementations from a flagship empirical management journal. Perhaps unsurprisingly, applications such as those in Appendix \ref{secA1} hardly disclose any equation. Typically, a complete maximum likelihood analysis is considered superfluous, and researchers turn to a handful of econometric sources for details. Some textbooks do the same with regard to scoring and information equations \citetext{\citealp[e.g.,][Ch.~8]{LongJ.Scott1997Rmfc}; \citealp[Ch.18]{Wooldridge}}.
	
	This article challenges the prevailing view that the computational structure of the Negative Binomial regression can be safely taken for granted. We examine widely-accepted results from routinely-cited resources and find problems in all of them, especially with regard to scoring and information equations in the so-called dispersion parameter. As a consequence we find no reliable expression for the element of the Fisher Information Matrix corresponding to such parameter. By elevating computations that are rarely made explicit the hoped for outcome is to remediate confusing notations that hinder comparison and conceal imprecision. 
	
	The remainder is structured as follows. The next section compares widely-accepted versions of the Negative Binomial specifications, while seeking verification by direct proof. Analysis is extended to include an unpublished approach implemented under the hood of a well-known, open-source software routine. Comparative insights are discussed an implemented in \texttt{R} with the aid of a simplified numerical example centered on clinical trial supply. A closing section summarizes key  practical implications of the proposed research, and its limitations.

	\section{Proposed approach}\label{section20}
	Even for elementary statistics concepts, such as quantiles and box plots, disagreement among alternative texts or software is not uncommon \citep[e.g.,][]{Langford01012006, Mcgill01021978}. Work in applied management research defers to high-quality econometrics textbooks for proofs and supporting evidence when problems arise in the application of existing methods \citep{Shang22}. But what if these trusted sources disagree? 
	
	To address this question for the Negative Binomial specification in count regression we set out to verify key results that are extensively relied upon in empirical management research and beyond---namely those in \cite{Cameron_Trivedi_2013}, \cite{Hilbe_2011}, and \cite{Lawless87}. We also consider an unpublished alternative implemented in the open-source language \texttt{R} \citep{R_core} by the pre-built routine \texttt{glm.nb} from the package \texttt{MASS} \citep{Venables.2002Masw}. In addition, we review equations that are disclosed in publicly available documentation for proprietary software such as \texttt{STATA} and \texttt{NCSS}.
	
	Tab.~\ref{tab:compare} summarizes upfront our comparative insights at each stage of a typical maximum likelihood analysis. What is clear from Tab.~\ref{tab:compare} is that specifications across all reviewed sources have limitations, although to a varying extent. The last column of Tab.~\ref{tab:compare} points to the striking absence of reliable results about the Fisher (or \textit{expected}) Information Matrix for parameters other than the regression coefficients.
	
	\vfill
	
	%\begin{adjustbox}{max width=\linewidth}	
	\begingroup	
	\renewcommand{\arraystretch}{0.6} % based on https://www.overleaf.com/latex/examples/latex-table-spacing-example/pwtdmxfmnbxc
	\begin{threeparttable}
		\caption{\label{tab:compare} Summary of comparative analysis. Symbols denote the presence of an attribute ($\bullet$) or whether the equations are correct (\ding{51}), incorrect (\ding{55}) or unspecified (n.s.)}
		\begin{tabular}{llccccccc}
			\toprule
			\multicolumn{2}{l}{References} & \multicolumn{3}{l}{Notation formalism} & \multicolumn{4}{l}{Stages of analysis\tnote{1}} \\
			\cmidrule(l{3pt}r{3pt}){3-5}\cmidrule(l{3pt}r{3pt}){6-9}
			& & \multicolumn{2}{l}{Parametization\tnote{2}} &  Gamma-free & LL & \multicolumn{2}{c}{SI} & FI \\                  
			\cmidrule(l{3pt}r{3pt}){3-4}\cmidrule(l{3pt}r{3pt}){7-8}
			& & DI & GS &   notation &  & SC & IN & \\		
			\midrule		
			\multicolumn{9}{l}{Academic references} \\
			& \cite{Cameron_Trivedi_2013}          & $\bullet$ &   & $\bullet$ & \ding{51} & \ding{51} & n.s. & \ding{55} \\		
			& \cite{Hilbe_2011}         & $\bullet$ &   &   & \ding{51}& \ding{55} & \ding{55}& n.s.\\               
			& \cite{Lawless87}         & $\bullet$ &   & $\bullet$   & \ding{55} & \ding{55}  & n.s. &\ding{55} \\ 	
			\multicolumn{9}{l}{Software} \\
			& \texttt{R MASS::glm.nb}\tnote{3} &           & $\bullet$   &   & \ding{51} & \ding{51} & \ding{51} & n.s.  \\				
			& STATA\tnote{4}  & $\bullet$& & &\ding{51}  & n.s.  & n.s. &	n.s. 	\\ 
			& NCSS\tnote{5}  & $\bullet$ &  &  $\bullet$  &\ding{51} &\ding{51} & \ding{55} &  \ding{55}		\\ 
			\bottomrule						
		\end{tabular}		
		\begin{tablenotes}[para]	
			\scriptsize
			\item[1] Following the structure of Sec.\ref{section2} LL: log-likelihood; SI: scoring (SC) and information (IN); FI: "expected" (Fisher) information.  
			\item[2] DI: dispersion parameter; GS: gamma scale parameter.
			\item[3] Based on function's scripts. \citet[][Ch.~11]{Venables.2002Masw} do not disclose the equations.
			\item[4] As reported on \url{https://www.stata.com/manuals/rnbreg.pdf};
			\item[5] As reported on \url{https://www.ncss.com/software/ncss/ncss-documentation/} \\
		\end{tablenotes}
	\end{threeparttable}
	%\end{adjustbox}
	\endgroup    
	\begingroup	
	\renewcommand{\arraystretch}{1}
	
	\endgroup
	
	It is not unusual to examine the limitations of widely-accepted, but potentially misleading results affecting the Negative Binomial specification \cite[e.g.,][]{SHONKWILER2016209}. Although developed in an adjacent context, we adopt the approach of \cite{lognormLCA}as a guiding framework for our work, as we engage in a comparative analysis with a view to: 
	
	\begin{itemize}
		\item Elevating computations that are often taken for granted, so to provide ``conversion steps'' between alternative notations that are rarely reconciled;
		\item Addressing potential issues in terms of (i) interoperability of software implementations; and (ii) interpretation of key analytical results from such implementations.
	\end{itemize}
	
	Against this backdrop, the next sections detail how the summary in Tab.~\ref{tab:compare} is arrived at.

	\section{Comparative analysis}\label{section2}   
\begin{figure}
		\centering
		\includegraphics[width=0.7\linewidth]{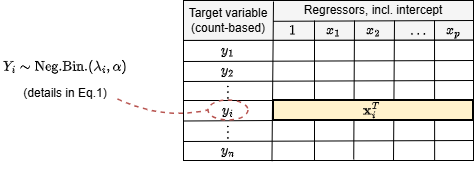}
		\caption{Schematic of a typical data layout for cross-section Negative Binomial regression.}
		\label{fig:fig-nbData}
	\end{figure}

	It is instructive to begin from the typical cross-section data layout for count regression  in Fig.~\ref{fig:fig-nbData}. An observed count $y_i \quad (i=1,2,..., n\in \mathbb{N})$  is taken to be the realization of a random variable $Y_i$ that is approximately Negative Binomial with conditional mean count $\lambda_i$. Such mean count may be related to a row vector $\mathbf{x}_i^T$ consisting of $p$ regressors and an intercept. 
	
	The Negative Binomial distribution mentioned in Fig.~\ref{fig:fig-nbData} can be thought of as a Poisson-Gamma mixture \citep[][Ch.5]{discrete_dist} in which the Poisson mean $\lambda_i$ is dependent on an ``unobservable disturbance'' that is Gamma-distributed with single (scale) parameter $\alpha$. Following \citet[][Ch.~18]{Greene2012} the resulting probability mass function (\textit{p.m.f.}) is:
	
	\begin{align}
		\mathrm{Pr}\left(Y_i=y_i | \mathbf{x_i}\right) \nonumber &=  \frac{\Gamma\left( y_i + \alpha \right)}{\Gamma\left( y_i + 1 \right) \Gamma\left( \alpha \right)}\left( \frac{\lambda_i}{\lambda_i + \alpha}\right)^{y_i}\left(\frac{\alpha}{\lambda_i + \alpha}\right)^\alpha \\
		& \equiv Y_i \sim \mathrm{Neg.Bin}\left(\lambda_i, \alpha \right) 
		\label{eq:negbinPmf}
	\end{align}
	
	where $\lambda_i,\alpha>0$ are the unknown parameters of the distributions involved in the mixture; $\Gamma(\cdot)$ is the gamma function; and $\theta=\alpha^{-1}$ is referred to as the \textit{dispersion} parameter. 
	
	To ensure positive expected counts, the function $\lambda_i=e^{\mathbf{x}_i^T\boldsymbol{\beta}}$ is typically chosen to relate the mean $\lambda_i$ of the distribution in Eq.~\eqref{eq:negbinPmf} to the vector $\mathbf{x}_i^T$ in Fig.~\ref{fig:fig-nbData} through the unknown vector $\boldsymbol{\beta}$ of regression coefficients, to be estimated simultaneously with $\alpha$---or its reciprocal $\theta$. 
	
	The familiar \textit{p.m.f.} in Eq.~\eqref{eq:negbinPmf} serves as starting point in many applications---from seminal work \cite[e.g.,][]{Hausman84} to more recent developments \cite[e.g.,][]{PlatformGrocery}.

	\subsection{Stage 1: Log-likelihood}\label{sec:gammafree}		
	As Tab.~\ref{tab:compare} suggests, confusions about the Negative Binomial specification begin as early as the log-likelihood equation---the first step in the method of maximum likelihood estimation. It is therefore worthwhile illustrating alternative notations that are widely used in practice.
	
	With the \textit{p.m.f.} in Eq.~\eqref{eq:negbinPmf} and the likelihood $\textrm{L}(\alpha) = \Pr\left(Y_1=y_1 \land Y_2=y_2 \land  \cdots \land Y_n=y_n \right)$, the log-likelihood $\ell\left(\alpha \right) = \ln \textrm{L}(\alpha) $ for the Negative Binomial specification may be written:
	
	\begin{align}
		\ell\left(\alpha \right) &=  \sum_{i=1}^n \left\{ \ln \frac{\Gamma(y_i+\alpha)}{\Gamma(\alpha)}-\ln\Gamma(y_i+1)   + y_i \ln e^{\mathbf{x}_i^T\boldsymbol{\beta}}  + \alpha\ln \alpha - (\alpha + y_i)\ln\left[\alpha + e^{\mathbf{x}_i^T\boldsymbol{\beta}} \right] \right\} \label{eq:NBllaG}
	\end{align}
	
	where $\alpha$ and $\boldsymbol{\beta}$ are parameters to be estimated. Invoking the following identity \citep[][p.~210]{Lawless87}: 
	
	\begin{equation}
		\frac{\Gamma\left(y_i + \alpha\right)}{\Gamma\left(\alpha\right)}=\prod_{j=0}^{y_i-1}\left( j+\alpha\right) \label{gammaprod}
	\end{equation}
	
	and substituting $\alpha=\theta^{-1}$ in Eq.~\eqref{eq:NBllaG} yields a ``gamma-free'' log-likelihood in the dispersion parameter  \citep[cf.][p.~81]{Cameron_Trivedi_2013}:
	
	\begin{equation}
		%\hspace*{-1.5cm} 
		\ell\left(\theta\right) = \sum_{i=1}^n \left\{ \sum_{j=0}^{y_i-1}\ln\left( j + \theta^{-1}  \right) - \ln\left( y_i! \right) + y_i\mathbf{x}_i^T\boldsymbol{\beta} +  y_i\ln\theta -\left( \theta^{-1} + y_i\right)\ln\left(1+\theta e^{\mathbf{x}_i^T\boldsymbol{\beta}}\right) \right\} \label{eq:maxlik2}
	\end{equation}
	
	An alternative formulation of Eq.~\eqref{eq:maxlik2} is given by \citet[][Ch.~8]{Hilbe_2011}, who retains the gamma function from the \textit{p.m.f.} in Eq.~\eqref{eq:negbinPmf}. A log-likelihood that features both the gamma function and the dispersion parameter is not so much an issue \textit{per se} as it is a fertile ground for imprecision when differentiating it---as we will show in the next sections.
	
	Comparing Eq.~\eqref{eq:maxlik2} with the  log-likelihood reported in \citet[][p.~210]{Lawless87} raises some concerns about the latter, and the results that rely on it. Yet these results have been  widely cited without being called into question \citetext{\citealp[e.g.,][]{Piegorsch90}; \citealp[Ch.~8]{LongJ.Scott1997Rmfc}; \citealp{Ismail07}; \citealp{NBtree}}. Tab.~\ref{tab:compare} flags this such discrepancy as non-trivial.

	\subsection{Stage 2: Scoring and information equations}\label{sec-scor}
	Standard maximum likelihood estimation requires that the log-likelihood examined in the previous section be differentiated with respect to (w.r.t.) each set of parameters---i.e., the regression coefficients $\boldsymbol{\beta}$ and the dispersion parameter $\theta$. The first case poses no particular challenge: all the sources examined here are in agreement on that regard. We report these results in Appendix \ref{sec:A2} for the sake of completeness, with a view to forming the elements of the Fisher Information Matrix in a later section. Yet discrepancies do arise when the log-likelihood is differentiated w.r.t. $\theta$, as demonstrated in the next sub-sections.
	
	\subsubsection{Discrepancies due to first-order conditions.}
	Differentiating the log-likelihood w.r.t. $\theta$ yields  \citep[cf.][Ch.3]{Cameron_Trivedi_2013}:				
	
	\begin{align}				
		\frac{\partial}{\partial \theta}\ell{\left(\boldsymbol{\beta},\theta\right)} 
		&=  \sum_{i=1}^n\left\{-\frac{1}{\theta^2}\sum_{j=0}^{y_i-1}\frac{1}{\left(j+\theta^{-1}\right)}+\frac{\ln{\left(1+\theta e^{\mathbf{x}_i^T\boldsymbol{\beta}}\right)}}{\theta^2}+\frac{y_i}{\theta}-\frac{\left(\theta^{-1}+y_i\right)e^{\mathbf{x}_i^T\boldsymbol{\beta}}}{1+\theta e^{\mathbf{x}_i^T\boldsymbol{\beta}}}\right\} \nonumber \\
		& = \sum_{i=1}^n\left\{\frac{1}{\theta^2}\left[-\sum_{j=0}^{y_i-1}\frac{1}{\left(j+\theta^{-1}\right)}+\ln{\left(1+\theta e^{\mathbf{x}_i^T\boldsymbol{\beta}}\right)}\right]+\frac{y_i- e^{\mathbf{x}_i^T\boldsymbol{\beta}}}{\theta\left(1+\theta e^{\mathbf{x}_i^T\boldsymbol{\beta}}\right)}\right\}
		%& = {g}_{\theta} 
		\label{eq:d1}
	\end{align}
	
	As Tab.~\ref{tab:compare} points out, there is disagreement between Eq.~\ref{eq:d1} and analogous results such as \citet[Eq.~8.14]{Hilbe_2011} and \citet[Eq.~2.4]{Lawless87}. In the latter case the discrepancy arises due to an incorrect log-likelihood equation, which we touched on earlier. The former case is more elusive, and requires further clarification.
	
	We seek to verify the scoring equations for the Negative Binomial specification provided in references such as \citet{Hilbe_2011, Hilbe2014}. To achieve that, we differentiate Eq.~\eqref{eq:NBllaG} instead of Eq.~\eqref{eq:maxlik2} after substituting $\alpha = \theta^{-1}$. We do so without attempting, for the moment, to find the derivative of the ratio between gamma functions that makes up the first term in Eq.~\eqref{eq:NBllaG}. 
	
	Comparing the expression thus obtained (not reported here) with Eq.\eqref{eq:d1}, and recalling the chain rule from calculus we obtain the identity:
	
	\begin{align}
		-\frac{1}{\theta^2}\sum_{j=0}^{y_i-1}\frac{1}{j+\theta^{-1}} &= \frac{\mathrm{\partial}}{\mathrm{\partial}\theta}\ln\left[\frac{\Gamma\left(y_i + \theta^{-1}\right)}{\Gamma\left(\theta^{-1}\right)} \right] \nonumber\\    
		& = -\frac{1}{\theta^2}\left\{ \frac{\mathrm{d}}{\mathrm{d}\alpha}\ln\left[\Gamma\left(y_i + \alpha\right)\right]-\frac{\mathrm{d}}{\mathrm{d}\alpha}\ln\left[\Gamma\left(\alpha\right)\right] \right\} \nonumber\\
		& = -\frac{1}{\theta^2}\left[\Psi\left(y_i + \alpha \right) - \Psi\left(\alpha \right) \right]  \label{eq:dlnGamma2}    
	\end{align}
	
	where $\Psi\left(\cdot\right)=\frac{\Gamma^{'}\left(\cdot\right)}{\Gamma\left(\cdot\right)}$ is the \textit{digamma function}. With Eq.~\eqref{eq:dlnGamma2} we examine \citet[][Eq.~8.14]{Hilbe_2011}---the equivalent of our Eq.~\eqref{eq:d1}---and notice that multiplication by $-\frac{1}{\theta^2}$ is omitted, thus undermining an otherwise correct result. While Eq.~\eqref{eq:dlnGamma2} is rarely stated explicitly in the literature, a better-known variation is obtained dividing both sides by $-\frac{1}{\theta^2}$ \citep[see e.g.,][Theorem 1]{Yu2024}. Familiarity with the latter result might lead to incorrect results even when a ``gamma-free'' notation is adopted \citep[e.g.,][Eq.~2.4]{Lawless87}. 
	
	So far, the ``gamma-free'' notation in  \citet[Ch.3]{Cameron_Trivedi_2013} appears more reliable. Yet notations that feature the digamma function are relevant in practical software implementation. As Tab.~\ref{tab:compare} suggests, this is the case for the pre-built Negative Binomial regression implementation \texttt{glm.nb} from the \texttt{R} package \texttt{MASS} \citep[Ch.~7]{Venables.2002Masw}, where the base-\texttt{R} command \texttt{digamma} returns $\Psi(\cdot)$. We will come back to this topic in Sec.~\ref{sec:discussion}.

	\subsubsection{Discrepancies due to second-order conditions.}
	The discrepancies between canonical references widen substantially when it comes the second derivative of the log-likelihood function in Eq.~\ref{eq:maxlik2}. Consistently with the previous sections, we begin by providing a ``gamma-free'' expression for such derivative:
	
	\begin{align}
		\frac{\partial^2}{\partial \theta^2}\ell\left(\boldsymbol{\beta},\theta \right)  & = \sum_{i=1}^{n}\left\{ \sum_{j=0}^{y_i-1} \frac{2j\theta+1}{\theta^2\left(j\theta+1\right)^2}+  \left[ \frac{ {\theta e^{\mathbf{x}_i^T\boldsymbol{\beta}}}{\left(1+\theta  e^{\mathbf{x}_i^T\boldsymbol{\beta}}\right)^{-1}} - 2\ln \left(1+\theta  e^{\mathbf{x}_i^T\boldsymbol{\beta}}\right) }{\theta^3} + \right.\right. \nonumber \\
		& \quad \left.\left.  - \left(y_i -  e^{\mathbf{x}_i^T\boldsymbol{\beta}} \right)\frac{\left(1+2\theta e^{\mathbf{x}_i^T\boldsymbol{\beta}}\right)}{\theta^2\left(1+\theta e^{\mathbf{x}_i^T\boldsymbol{\beta}}\right)^2} \right] \right\} \nonumber \\
		&= \sum_{i=1}^{n}\left\{ \frac{1}{\theta^2} \sum_{j=0}^{y_i-1} \frac{\theta\left(2j+\theta^{-1}\right)}{\theta^2\left(j+\theta^{-1}\right)^2} + \frac{1}{\theta^2}\left[ \frac{\theta e^{\mathbf{x}_i^T\boldsymbol{\beta}}-\left(1+\theta e^{\mathbf{x}_i^T\boldsymbol{\beta}}\right)2\ln\left(1+\theta e^{\mathbf{x}_i^T\boldsymbol{\beta}}\right)}{\theta\left(1+\theta e^{\mathbf{x}_i^T\boldsymbol{\beta}}\right)} + \right.\right.  \nonumber \\
		&\quad -\left. \left.\frac{\left(y_i- e^{\mathbf{x}_i^T\boldsymbol{\beta}}\right)\left(1+2\theta e^{\mathbf{x}_i^T\boldsymbol{\beta}}\right)}{\left(1+\theta e^{\mathbf{x}_i^T\boldsymbol{\beta}}\right)^2} \right]  \right\} \nonumber  \\
		&=  \sum_{i=1}^{n}\left\{ \frac{1}{\theta^3} \sum_{j=0}^{y_i-1} \frac{2j+\theta^{-1}}{\left(j+\theta^{-1}\right)^2} - \frac{1}{\theta^3}\left[\frac{\theta\left(1+2\theta e^{\mathbf{x}_i^T\boldsymbol{\beta}}\right)\left(y_i- e^{\mathbf{x}_i^T\boldsymbol{\beta}}\right)}{\left(1+\theta e^{\mathbf{x}_i^T\boldsymbol{\beta}}\right)^2}+ \right.\right.\nonumber  \\
		& \quad \left.\left. - \frac{\theta e^{\mathbf{x}_i^T\boldsymbol{\beta}}\left(1+\theta e^{\mathbf{x}_i^T\boldsymbol{\beta}}\right)}{\left(1+\theta e^{\mathbf{x}_i^T\boldsymbol{\beta}}\right)^2} + 2\ln\left(1+\theta e^{\mathbf{x}_i^T\boldsymbol{\beta}}\right) \right]  \right\}\label{eq:d2}
	\end{align}
	
	Eq.~\eqref{eq:d2} hardly resembles analogous ``gamma-free'' results \cite[e.g.,][]{Lawless87}, suggesting non-trivial discrepancies. One deduces from \citet[][Eq.~3.32]{Cameron_Trivedi_2013} that the derivative assumed in that work, while undisclosed, differs from ours, too. By contrast, \citet[][Eq.~8.18]{Hilbe_2011} bears greater resemblance with Eq.~\eqref{eq:d2}, although the presence of the gamma function and its derivatives conceals some imprecision, as we demonstrate next.
	
	We seek to demonstrate that, despite the apparent similarity, it is not the case that \citet[][Eq.~8.18]{Hilbe_2011} is equivalent to our Eq.~\eqref{eq:d2}. As in the previous section, we differentiate Eq.~\eqref{eq:NBllaG} twice after substituting $\alpha = \theta^{-1}$ without attempting to find the derivative of the ratio between gamma functions in said equation. Comparing the expression thus obtained (not showed here) with Eq.~\eqref{eq:d2} yields the following equivalence:
	
	\begin{align}
		%\hspace*{-2cm}
		\frac{1}{\theta^3}\sum_{j=0}^{y_i-1}\frac{2j+\theta^{-1}}{\left(j+\theta^{-1} \right)^2} &= \frac{\mathrm{d}^2}{\mathrm{d}\theta^2}\ln\left[\frac{\Gamma\left(y_i+\theta^{-1}\right)}{\Gamma\left(\theta^{-1}\right)}\right] \nonumber \\
		& = -\frac{1}{\theta^2}\left\{\frac{\mathrm{d}^2}{\mathrm{d}\alpha^2}\ln\left[\frac{\Gamma\left(y_i+\alpha\right)}{\Gamma\left(\alpha\right)}\right]\right\} \nonumber \\   
		&= \frac{2}{\theta^3} \left[\Psi\left(y_i + \alpha \right) - \Psi\left(\alpha \right) \right] - \frac{1}{\theta^4}\left[\Psi^{'}\left({y_i +\alpha}\right) - \Psi^{'}\left({\alpha}\right)\right] \label{eq:NBtrigam}
	\end{align}
	
	where  $\Psi^{'}\left( \cdot \right)$  is  the \textit{trigramma} function i.e., the derivative of the \textit{digamma} function introduced previously. Eq.~\eqref{eq:NBtrigam} suggests that the portion of the second derivative in \citet[][Eq.~8.18]{Hilbe_2011} which is expressed in terms of the trigamma function is incorrect. We speculate that, once again, this might be due to the fact that, while Eq.~\eqref{eq:NBtrigam} is rarely encountered in the literature, one of its elements is a relatively well-known identity, namely \citep[cf.][]{Yu2024}:
	
	\begin{align}
		\Psi^{'}\left({y_i + \alpha}\right) - \Psi^{'}\left({\alpha}\right) & =  \frac{d}{d\alpha}\left[\Psi\left(y_i + \alpha \right) - \Psi\left(\alpha \right)\right] \nonumber \\
		& = \sum_{j=0}^{y_i-1}\frac{d}{d\alpha}\frac{1}{j+\alpha} \nonumber \\
		& = -\sum_{j=0}^{y_i-1}\frac{1}{\left(j+\alpha\right)^2} \label{eq:triga2}
	\end{align}
	
	It is not implausible that,	contrary to what Eq.~\ref{eq:NBtrigam}  suggests, some published results \citetext{\citealp[e.g.,][Eq.~8.18]{Hilbe_2011}; \citealp[Eq.~2.6 and 2.8]{Lawless87}} might be based on the incorrect assumption that $\frac{\mathrm{d}^2}{\mathrm{d}\theta^2}\ln\left[\frac{\Gamma\left(y_i+\theta^{-1}\right)}{\Gamma\left(\theta^{-1}\right)}\right]$ on the right-hand side of Eq.~\ref{eq:NBtrigam} is identical to $\Psi^{'}\left({y_i + \alpha}\right) - \Psi^{'}\left({\alpha}\right)$ on the left-hand side of Eq.~\ref{eq:triga2} or, equivalently, to $-\sum_{j=0}^{y_i-1}\frac{1}{\left(j+\alpha\right)^2}$ on its right-hand side. 
	
	\begin{remark}
		Although $\alpha$ and $\theta^{-1}$ may be used interchangeably, we find that rarely is the chain rule applied correctly. Often, the work examined invokes results about the gamma function's derivatives such as Eq.~\ref{eq:triga2} and substitutes $\alpha=\theta^{-1}$, which appears to be incorrect. 
	\end{remark}

	\subsection{Stage 3: Fisher Information Matrix}
	The issues encountered in the previous sections have a knock-on effect on the Fisher or ``expected'' information matrix. This aspect of the Negative Binomial specification is, surprisingly overlooked: the examined sources either avoid engaging with it or, if they do, they arrive at unreliable results expressed in a notation that is, at best, impractical to implement. In the next sub-section we seek to address this methodological gap.
	
	\subsubsection{Context}
	The Fisher or ``expected'' information matrix is defined as the expectations of minus the second derivatives of the log-likelihood function \citep[e.g.,][p.~211]{Lawless87}. Evaluating the information matrix at the maximum likelihood estimators (MLE) for the parameters $\boldsymbol{\beta}$ and $\alpha$, once these have been obtained, is a necessary step to obtain the asymptotic covariance matrix of these estimators and, ultimately, test their significance \citep[][p.~561]{Greene2012}.
	
	For the Negative Binomial specification, and given some widely-agreed results reported for completeness in Appendix~\ref{sec:A2}, the Fisher information matrix is known to have the following block-diagonal structure \citep[cf.][Ch.3]{Cameron_Trivedi_2013}:
	
	\begin{equation}
		\mathbf{I}({\boldsymbol{\beta}}, {\theta}) =  \begin{bmatrix} -\textrm{E} \left( \frac{\partial^2}{\partial \boldsymbol{\beta}\partial \boldsymbol{\beta}^T}\ell{\left(\boldsymbol{\beta},\theta\right)} \right) & \mathbf{0}  \\ \mathbf{0} & -\textrm{E} \left(\frac{\partial^2}{\partial \theta^2}\ell\left(\boldsymbol{\beta},\theta \right)\right) \end{bmatrix}     \label{eq:fiMatgen}
	\end{equation}
	
	with $-\textrm{E} \left( \frac{\partial^2}{\partial \boldsymbol{\beta}\partial \boldsymbol{\beta}^T}\ell{\left(\boldsymbol{\beta},\theta\right)} \right)=\sum_{i=1}^n{\frac{e^{\mathbf{x}_i^T\boldsymbol{\beta}}}{1+\theta e^{\mathbf{x}_i^T\boldsymbol{\beta}}}\mathbf{x}_i\mathbf{x}_i^T}$ as per Eq.~\eqref{eq:fibeta} in the Appendix.
	
	Yet arriving at the complete Fisher Information Matrix in Eq.~\eqref{eq:fiMatgen} is no trivial task due to the element $-\textrm{E} \left(\frac{\partial^2}{\partial \theta^2}\ell\left(\boldsymbol{\beta},\theta \right)\right)$: this is due not only to disagreeing results about the underlying second derivative, as discussed in the previous section; but also to the expectation operation, which poses computational challenges in its own right and may require approximation strategies \cite[see e.g.,][]{Yu2024}. These issues are addressed next.
	
	\subsubsection{Expectations}
	In the absence of reliable guidance from the literature, we seek to obtain a reliable expression for the bottom-right element in Eq.~\eqref{eq:fiMatgen}. To ease the notation we write $\lambda_i=e^{\mathbf{x}_i^T\boldsymbol{\beta}}$; and $\Pr\left(Y_i=y_i\right)$ is the Negative Binomial \textit{p.m.f.} in Eq.~\eqref{eq:negbinPmf}. We also recall that the distribution mean is $\mathrm{E}\left[Y_i\right]=\sum_{y_i=0}^{\infty} y_i\mathrm{Pr}\left(Y_i=y_i\right)=\lambda_i$. The expected value of interest is:	
	
	\begin{align}
		\textrm{E}\left[-\frac{\partial^2\ell{\left(\boldsymbol{\beta},\theta\right)}}{\partial\theta^2}\right] &= -\sum_{y_i=0}^{\infty}\sum_{i=1}^{n}\left\{ \frac{1}{\theta^3} \sum_{j=0}^{y_i-1} \frac{2j+\theta^{-1}}{\left(j+\theta^{-1}\right)^2} - \frac{1}{\theta^3}\left[\frac{\theta\left(1+2\theta \lambda_i\right)\left(y_i- \lambda_i\right)}{\left(1+\theta \lambda_i\right)^2}+ \right.\right. \nonumber\\
		& \quad \left.\left. - \frac{\theta \lambda_i\left(1+\theta \lambda_i\right)}{\left(1+\theta \lambda_i\right)^2} + 2\ln\left(1+\theta \lambda_i\right) \right]  \right\}\Pr\left(Y_i=y_i\right) \nonumber\\
		& = \frac{1}{\theta^3} \sum_{i=1}^{n} \left[\frac{ 2\left(1+\theta\lambda_i \right)^2 \ln\left(1+\theta\lambda_i \right)}{\left(1+\theta\lambda_i \right)^2 } - \frac{\theta\lambda_i\left(1+\theta\lambda_i \right)+ \theta\lambda_i\left(1+2\theta\lambda_i\right)}{\left(1+\theta\lambda_i \right)^2} + \right.  \nonumber\\
		& \quad - \left. \sum_{y_i=0}^{\infty} \sum_{j=0}^{y_i-1} \frac{2j+\theta^{-1}}{\left(j+\theta^{-1}\right)^2}\mathrm{Pr}\left(Y=y_i\right)  \right] + \frac{1}{\theta^3}\sum_{i=1}^n\theta\frac{\left(1+2\theta\lambda_i\right)}{\left(1+\theta\lambda_i\right)^2}\mathrm{E}\left[y_i\right] \nonumber \\
		&=\frac{1}{\theta^3}\sum_{i=1}^n\left[2\ln\left(1+\theta\lambda_i\right) - \frac{\theta\lambda_i}{1+\theta\lambda_i}-\sum_{y_i=0}^{\infty} \sum_{j=0}^{y_i-1} \frac{2j+\theta^{-1}}{\left(j+\theta^{-1}\right)^2} \Pr\left(Y_i=y_i\right) \right] 
		\label{eq:fish} 
	\end{align}

	Eq.~\ref{eq:fish} differs considerably from alternative results available in the literature examined. These, in turn, bear little resemblance with each other. Some references avoid computing the expected information altogether, and focus on the ``observed`` information, instead \citep[e.g.,][Ch.8]{Hilbe_2011}. Other references engage with the relevant computations---without disclosing them---but arrive at erroneous results \citep[e.g.,][Eq.~2.8]{Lawless87}. 
	
	The result in \citet[Eq.~3.32]{Cameron_Trivedi_2013} is particularly confusing, and worth discussing further. Although never explicitly, this source invokes Eq.~\ref{eq:triga2} in a context where the chosen parametrization is $\theta$, not $\alpha$. For reasons mentioned  earlier, this appears to be incorrect. What is more, the notation in \citet[Eq.~3.32]{Cameron_Trivedi_2013} suggests that the required expectation has been evaluated, yet the reported result is an evaluation of the ``observed'', not the ``expected'' information matrix.	
	
	\subsubsection{Approximations}
	Even if ``correct'' vis-\`{a}-vis the alternatives examined so far, Eq.~\eqref{eq:fish} is cumbersome and it features an infinite sum that does not lend itself to practical computations. Approximations have been recently proposed \citep[e.g.,][]{Yu2024} but assume a parametrization in $\alpha$ as opposed to the dispersion parameter $\theta$ that is prevalent in the literature.
	
	To take advantage of such approximation we seek to obtain $\mathrm{E}\left[-\frac{\partial^2}{\partial \alpha^2}\ell\left(\boldsymbol{\beta},\alpha \right)\right]$. With the log-likelihood in Eq.~\ref{eq:NBllaG} we re-state results in Sec.~\ref{sec-scor} in the alternative parameterization $\alpha$:
	
	\begin{itemize}		
		\item The first derivative of Eq.~\ref{eq:NBllaG} w.r.t. $\alpha$ is as follows:		
		\begin{align}
			\frac{\partial}{\partial \alpha}\ell\left(\boldsymbol{\beta},\alpha \right) &= \sum_{i=1}^{n}\left\{ \left[ \Psi\left(y_i + \alpha \right) - \Psi\left(\alpha\right)\right] + \ln\alpha + 1 - \ln\left(\lambda_i + \alpha\right) - \frac{\alpha + y_i}{\lambda_i + \alpha}  \right\} \nonumber \\
			& = \sum_{i=1}^{n}\left\{ \sum_{j=0}^{y_i - 1}\frac{1}{j+\alpha}-\ln\left(1+\alpha^{-1}\lambda_i\right) - \frac{y_i-\lambda_i}{\alpha\left(1+\alpha^{-1}\lambda_i\right)} \right\}
			\label{eq:d1_alpha}
		\end{align}		
		\noindent Eq.~\ref{eq:d1_alpha} is the counterpart of Eq.~\ref{eq:d1}. The first line is implemented under the hood by \texttt{MASS::glm.nb} with the \texttt{R} command \texttt{digamma} returning the digamma function $\Psi\left(\cdot\right)$. The identity $ \Psi\left(y_i + \alpha \right) - \Psi\left(\alpha\right)= \sum_{j=0}^{y_i - 1}\frac{1}{j+\alpha}$ follows form Eq.~\ref{eq:dlnGamma2}. 
		\item The second derivative of Eq.~\ref{eq:NBllaG} w.r.t. $\alpha$ is:
		\begin{align}
			%\hspace*{-0.5cm} 
			\frac{\partial^2}{\partial \alpha^2}\ell\left(\boldsymbol{\beta},\alpha \right) & = \sum_{i=1}^{n}\left\{\Psi^{'}\left(y_i + \alpha\right) - \Psi^{'}\left(\alpha\right) + \frac{1}{\alpha} - \frac{2}{\lambda_i + \alpha} + \frac{\alpha + y_i}{\left(\lambda_i + \alpha\right)^2} \right\} \nonumber \\
			& = \sum_{i=1}^{n}\left\{ -\sum_{j=0}^{y_i - 1}\frac{1}{\left(j+\alpha\right)^2} + \frac{\lambda_i^2+\alpha y_i}{\alpha\left(\alpha + \lambda_i \right)^2}\right\} 
			\label{eq:d2_alpha}
		\end{align} 
		\noindent Eq.~\ref{eq:d2_alpha} is the counterpart of Eq.~\ref{eq:d2}. The first line is implemented under the hood by \texttt{MASS::glm.nb} with the \texttt{R} command \texttt{trigamma} returning the trigamma function $\Psi^{'}\left(\cdot\right)$. The identity $\Psi^{'}\left(y_i + \alpha\right) - \Psi^{'}\left(\alpha\right) = -\sum_{j=0}^{y_i - 1}\frac{1}{\left(j+\alpha\right)^2}$ corresponds to Eq.~\ref{eq:triga2}.
		\item Taking the expectation of minus the second derivative in Eq.~\eqref{eq:d2_alpha} yields the sought-after element of the Fisher information matrix for the parametrization in $\alpha$:    
		\begin{align}
			%\hspace*{-0.5cm} 
			\mathrm{E}\left[-\frac{\partial^2}{\partial \alpha^2}\ell\left(\boldsymbol{\beta},\alpha \right)\right] &= - \sum_{i=1}^{n} \left\{ \mathrm{E}\left[\Psi^{'}\left(y_i + \alpha\right) \right] -\Psi^{'}\left(\alpha\right) + \frac{1}{\alpha} - \frac{2}{\lambda_i + \alpha} + \frac{\alpha + \mathrm{E}\left[Y_i\right]}{\left(\lambda_i + \alpha\right)^2} \right\} \nonumber  \\
			&= \sum_i^n \left\{ \left[\sum_{y_i = 0}^\infty \mathrm{Pr}\left(Y_i=y_i\right)  \sum_{j=0}^{y_i - 1}\frac{1}{\left(j+\alpha\right)^2}\right] -\frac{1}{\alpha} + \frac{1}{\alpha + \lambda_i} \right\}  \nonumber \\
			&= \sum_i^n \left\{-\frac{\alpha^{-1}\lambda_i}{\alpha\left(1 + \alpha^{-1}\lambda_i\right)} + \sum_{j = 0}^\infty \frac{\Pr\left(Y_i>j\right) }{\left(j+\alpha\right)^2} \right\}
			\label{eq:info_alpha}
		\end{align}
		\noindent where $\mathrm{E}\left[Y_i\right]=\lambda_i$ is the expected value of the random count-based variable $Y_i$; and the identity $\textrm{E}\left[\Psi_1\left(Y_i+\alpha\right)\right]-\Psi_1\left(\alpha\right) =-\sum_{j=0}^{\infty}{\frac{\Pr{\left(Y_i>j\right)}}{\left(j+\alpha\right)^2}}$ is based on  \citet[][Theorem 1]{Yu2024}, although a similar expression may  be found in \citet[p.211]{Lawless87}.
	\end{itemize}
	
	Not only does Eq.~\eqref{eq:info_alpha} look less cumbersome than Eq.~\eqref{eq:fish}---its counterpart in the dispersion parameter $\theta$; but also the infinite sum $\sum_{j = 0}^\infty \frac{\mathrm{Pr}\left(Y_i>j\right)}{\left(j+\alpha\right)^2}$ can be bounded by some integer $M>0$ with known error.  Specifically, invoking \citet[][Theorem 2]{Yu2024} we write:
	
	\begin{equation}
		\mathrm{E}\left[-\frac{\partial^2}{\partial \alpha^2}\ell\left(\boldsymbol{\beta},\alpha \right)\right] \approx \sum_{i=1}^n \left[\sum_{j = 0}^M \frac{1- F_i(j)}{\left(j+{\alpha}\right)^2} -\frac{e^{\mathbf{x}_i^T{\boldsymbol{\beta}}}}{\alpha^2 + \alpha e^{\mathbf{x}_i^T \boldsymbol{\beta}}}\right]\label{eq:approxFIalpha}
	\end{equation}
	
	where $F_i(j)=\Pr(Y_i\leq j| \lambda_i, \alpha)$ is the Cumulative Distribution Function (CDF) given the parametrization in Eq.~\eqref{eq:negbinPmf}; and the approximation error of truncating the infinite sum  at a pre-specified integer $M>0$ is bounded as follows \citep[cf.][Theorem 2]{Yu2024}:
	
	\begin{equation}
		\sum_{j = 0}^\infty \frac{1-F_i(j)}{\left(j+\alpha\right)^2} - \sum_{j = 0}^M \frac{1-F_i(j)}{\left(j+\alpha\right)^2}\leq \frac{1-F_i(M+1) }{\alpha+M}
	\end{equation}
	
	with opportunities for further reduction of such error, which are beyond scope here. 
	
	For the sake of completeness we seek an equivalent of Eq.~\eqref{eq:approxFIalpha} in the parameter $\theta$. Beginning from the latter and substituting Eq.~\eqref{eq:NBtrigam} allows us to (i) invoke the theorems about the expectation of the digamma and trigamma functions in \citet{Yu2024} and (ii)  where there is no risk of confusion, substitute $\alpha=\theta^{-1}$. The result is as follows:
	
	\begin{align}
		\textrm{E}\left[-\frac{\partial^2\ell{\left(\boldsymbol{\beta},\theta\right)}}{\partial\theta^2}\right] 
		%& = \sum_{i=1}^n\left\{\frac{1}{\theta^3}\left[2\ln\left(1+\theta\lambda_i\right) - \frac{\theta\lambda_i}{1+\theta\lambda_i}\right]-\frac{2}{\theta^3}\left[\textrm{E}\left[\Psi\left(Y_i + \alpha \right)\right] - \Psi\left(\alpha \right) \right]  + \right. \nonumber \\
		%&\quad +\left. \frac{1}{\theta^4} \left[\textrm{E}\left[\Psi^{'}\left({Y_i +\alpha}\right)\right] - \Psi^{'}\left({\alpha}\right)\right] \right\} \nonumber \\
		& = \sum_{i=1}^n\left\{\frac{1}{\theta^3}\left[2\ln\left(1+\theta\lambda_i\right) - \frac{\theta\lambda_i}{1+\theta\lambda_i}\right]-\frac{2}{\theta^3} \sum_{j=0}^{\infty}\frac{\Pr(Y_i>j)}{\theta^{-1}+j}  - \frac{1}{\theta^4}\sum_{j=0}^{\infty}\frac{\Pr(Y_i>j)}{(\theta^{-1}+j)^2} \right\} \nonumber \\
		& =\frac{2}{\theta^3}\sum_{i=1}^n \left[ \ln\left(1+\theta\lambda_i\right) - \sum_{j=0}^{\infty}\frac{\Pr(Y_i>j)}{\theta^{-1}+j} \right]  + \frac{1}{\theta^4}\sum_{i=1}^n \left\{\sum_{j=0}^{\infty}\frac{\Pr(Y_i>j)}{(\theta^{-1}+j)^2} - \frac{\theta\lambda_i}{\theta^{-1}(1+\theta\lambda_i)}\right\} \nonumber \\
		& \approx \frac{2}{\theta^3} \sum_{i=1}^n\left[ \ln\left(1+\theta  e^{\mathbf{x}_i^T\hat{\boldsymbol{\beta}}}\right) - \sum_{j=0}^{M}\frac{1-F_i(j)}{\theta^{-1}+j} \right] + \frac{1}{\theta^4}\textrm{E}\left[-\frac{\partial^2\ell{\left(\boldsymbol{\beta},\alpha\right)}}{\partial\alpha^2}\right] 
		\label{eq:ApproxfisherInfoTheta}
	\end{align}
	
	where $\textrm{E}\left[-\frac{\partial^2\ell{\left(\boldsymbol{\beta},\alpha\right)}}{\partial\alpha^2}\right]$ refers to the approximation in Eq.~\eqref{eq:approxFIalpha}; and $F_i(\cdot)$ is the Negative Binomial CDF mentioned previously. Eq.~\eqref{eq:ApproxfisherInfoTheta} provides a computable approximation for Eq.~\eqref{eq:fish} in a notation coherent with the Fisher information matrix in Eq.~\eqref{eq:fiMatgen}.
	
	\section{Discussion}\label{sec:discussion}
	Throughout Sec.~\ref{section2} our aim has been to substantiate the comparative insights summarized in Tab.~\ref{tab:compare}.  In the next sub-sections we discuss our findings and illustrate them, for practical relevance, with the aid of a simplified numerical example centred on clinical trial supply. 
	
	\subsection{Summary of findings} 
	The comparative analysis detailed in the previous section may be summarized as follows:	
	
	\begin{enumerate}
		\item Scoring and information equations in the dispersion parameter of the Negative Binomial specification are generally overlooked, with disagreement between sources widening as soon as the log-likelihood function is differentiated w.r.t. $\theta$ as opposed to the familiar regression coefficients $\boldsymbol{\beta}$. 
		\item Notations centered on the dispersion parameter $\theta$ often feature the gamma function and its derivatives, which take as argument the reciprocal $\alpha=\theta^{-1}$. While not incorrect \textit{per se} this combination appears conducive to incorrect results about the scoring and information equations related to $\theta$.
		\item We were unable to find a reliable expression for the element of the Fisher information matrix corresponding to the dispersion parameter in the sources examined. As a consequence, these sources provide no guidance on how to engage with issues concerning the need for computable approximations of such element e.g.,  Eq.~\eqref{eq:approxFIalpha}.         
	\end{enumerate}			
	
	A practical insight emerging from our findings is that most issues highlighted above---with the exception of those related to evaluating the expectations in the Fisher information matrix---are currently better addressed under the hood of open-source software routines in \texttt{R}, than they are in routinely-cited references on Negative Binomial regression.  
	
	Specifically, the function \texttt{glm.nb} from the \texttt{MASS} library \citep[Ch.~7]{Venables.2002Masw} mentioned in  Tab.~\ref{tab:compare} relies consistently on the parameter $\alpha$. In Sec.~\ref{sec-scor} we find that such parameterization, as implemented in \texttt{glm.nb}, is (i) conducive to a less cumbersome or ambiguous notation when formulating scoring and information equations in the parameter $\alpha$, as shown in Eq.~\eqref{eq:d1_alpha}-\eqref{eq:info_alpha}; and (ii) convenient for the evaluation of expectations in the Fisher information matrix, as shown in Eq.~\eqref{eq:approxFIalpha}.
	
	Yet equations formulated in the dispersion parameter $\theta$ remain overwhelmingly popular, whereas the equations implemented under the hood by \texttt{MASS::glm.nb} are unpublished and only accessible by engaging with unannotated scripts:  \citet{Venables.2002Masw} do not disclose the equations; although work detailing this function is mentioned in a footnote in \citet{GardnerWilliam1995RAoC}, it does not appear to be available in the public domain.
	
	Confusion may arise, especially for \texttt{R} users.  For instance \cite{NBtree} defer to both \citet{Lawless87} and \cite{Venables.2002Masw}: yet the former engages with the expected information matrix, and returns results that seem incorrect; the latter avoids computing expectations altogether, opting for an ``observed'' information approach, instead.
	
	For the sake of clarity we provide as Online Supplement \texttt{R} scripts that implement a ``modified'' version of \texttt{MASS::glm.nb}. We keep the original's parametrization in $\alpha$ and alternating estimation procedure, but introduce the following modification: the variance of the MLE $\hat{\alpha}$ is computed by approximation as the inverse of the ``expected'' information matrix in Eq.~\eqref{eq:approxFIalpha}---evaluated at the MLE of the Negative Binomial parameters:
	
	\begin{equation}\label{eq:varalpha}
		\widehat{\textrm{Var}}(\hat{\alpha}) = \left\{\mathrm{E}\left[-\frac{\partial^2}{\partial \alpha^2}\ell\left(\boldsymbol{\beta},\alpha \right)\right]_{\boldsymbol{\beta}=\hat{\boldsymbol{\beta}},\alpha=\hat{\alpha}}\right\}^{-1}  \approx \left\{ \sum_{i=1}^n \left[\sum_{j = 0}^M \frac{1-\hat{F}_i(j)}{\left(j+\hat{\alpha}\right)^2} -\frac{e^{\mathbf{x}_i^T\hat{\boldsymbol{\beta}}}}{\hat{\alpha}^2 + e^{\mathbf{x}_i^T\hat{\boldsymbol{\beta}}}}\right] \right\}^{-1} 
	\end{equation}
	
	The squared root of the result in Eq.~\eqref{eq:varalpha} is the estimator's standard error.
	
	We close this sub-section touching on the relevance of our findings for modelling extensions such as rate-parametrization and zero-truncation. In the former case the count-based variable may be expressed as a rate relative to e.g., a time duration. The latter extension is useful when zero counts are missing from the observed data. For the Negative Binomial specification, work dedicated to these extensions typically omits scoring and information equations in the dispersion parameter, or its reciprocal \citep[see e.g.,][respectively]{zwilling, GroggerJ.T.1991Mftc}. For zero-truncated Negative Binomial regression, ``Stage 2'' insights from \ref{sec-scor} have direct relevance, as the form of the derivatives of the log-likelihood function includes results from the non-truncated case \cite[see e.g.,][]{GURMU1992347} .
	
	\subsection{Numerical example}\label{sec-example}
	For practical relevance we introduce a simplified numerical example in Appendix~\ref{sec:A3} that follows the layout in Tab.~\ref{tab:compare}, with one count-based variable (DSR) serving as target.
	
	Although simplified, the data in Appendix~\ref{sec:A3} are a small, pseudonomyzed and obfuscated excerpt taken from a dataset---which remains confidential---on clinical supply disruptions.
	
	Although it is not the aim of this paper to provide industry insights based on a simplified data excerpt. we feel that more context may add to the practical relevance of our work.  The supply of investigational medicinal products (IMP) to  patients  enrolled in clinical studies is characterized by  high service level requirements, and shortfall is heavily penalized  \citetext{\citealp[see e.g.,][Ch.~14]{Mills}; \citealp[]{geyman_settanni_srai_2020}}. Yet disruptions in pharmaceutical supply chains has received little attention in empirical operations management. For instance only one reference in Sec.~\ref{secA1} deals with medicines recalls \citep{JOM11_Ball}. Work on the empirical analysis of clinical supply disruption is scant \citetext{\citealp[e.g.,][]{Settanni.2026}, \citealp[]{Anisimov.2009}}. 
	
	Against this backdrop, we seek to compare alternative estimates for the parameter $\alpha$ and its standard error obtained by fitting a Negative Regression model to the data in Tab.~\ref{tab:tab2} with the following options implemented in \texttt{R}: (i) our proposed implementation based on Eqs.~\eqref{eq:NBllaG}, \eqref{eq:d1_alpha}---\eqref{eq:approxFIalpha} and \eqref{eq:varalpha} (provided as Online Supplement); (ii) the pre-built functions \texttt{glm.nb} from the package \texttt{MASS} \citep{Venables.2002Masw}; and (iii) the pre-built function \texttt{gamlss} from the package \texttt{GAMLSS} \citep{GAMLSS} with input argument \texttt{family = "NBI"}. It is worth emphasising that \texttt{gamlss} is based on Generalized Additive Models as the underlying statistical framework---which differes from the Generalized Linear Models implemented by \texttt{glm.nb}. The result is shown in Tab.~\ref{tab:tab4}.
	
	\begingroup	
	\renewcommand{\arraystretch}{0.6}       
	\resizebox{\ifdim\width>\linewidth\linewidth\else\width\fi}{!}{
		\begin{threeparttable}	        
			\caption{Streamlined numerical example, clinical trial supply (SE: standard errors)}\label{tab:tab4}         
			\begin{tabular}[t]{lrrrrrrr}
				\toprule
				\multicolumn{1}{l}{ } & \multicolumn{3}{l}{Own implementation} & \multicolumn{2}{l}{\texttt{glm.nb}} & \multicolumn{2}{l}{\texttt{glmss}} \\
				\cmidrule(l{3pt}r{3pt}){2-4} \cmidrule(l{3pt}r{3pt}){5-6} \cmidrule(l{3pt}r{3pt}){7-8}
				& Estimate & SE expected & SE observed & Estimate & SE & Estimate & SE\tnote{1}\\
				\midrule
				intercept & -0.92688 & 1.12779 & 1.16034 & -0.92688 & 1.12779 & -0.92634 & 1.39209\\
				BIO & 0.15436 & 1.03224 & 1.05444 & 0.15436 & 1.03224 & 0.15422 & 1.36541\\
				DUR & 0.02627 & 0.01167 & 0.01256 & 0.02627 & 0.01167 & 0.02625 & 0.01377\\
				CLI & -0.00355 & 0.00530 & 0.00529 & -0.00355 & 0.00530 & -0.00355 & 0.00733\\
				SUB & 0.00168 & 0.00122 & 0.00120 & 0.00168 & 0.00122 & 0.00168 & 0.00188\\
				$\alpha$ & 6.05464 & 7.16712 & 6.36378 & 6.05464 & 6.36378 & 5.94385 & {n.a.}\tnote{2}\\
				\bottomrule
			\end{tabular}
			\begin{tablenotes}[para]
				\scriptsize	
				\item[1] For an object \texttt{obj\_g} fitted with \texttt{gamlss} the following computations are required: SE obtained as the square root of the appropriate elements of \texttt{diag(vcov(obj\_g))}; $\alpha$ requires following computation: \texttt{1/exp(obj\_g\$sigma.coefficients)}.
				\item[2] Based on the above the SE returned by \texttt{gamlss} refers to $\ln \theta$ and, as such, is not reported here. 
			\end{tablenotes}
		\end{threeparttable}        
	}        
	\endgroup
	\begingroup	
	\renewcommand{\arraystretch}{1}
	\endgroup

	The alternatives examined in Tab.~\ref{tab:tab4} hardly differ for their estimates of the regression parameters, reflecting the agreement in the literature. Yet \texttt{glm.nb} adopts a ``hybrid'' approach towards the Fisher information matrix: the standard errors for the regression coefficients are underpinned by the ``expected'' information in Eq.~\ref{eq:fibeta}, whereas the standard error of $\alpha$ is underpinned by ``observed'' information. Unlike the alternatives, our implementation evaluates the expectations in the Fisher information matrix for $\alpha$, handling the necessary approximations, thus making a clear distinction between expected and observed information. 
	
	\section{Closing remarks}	
	This work reexamines results from a selection of routinely-cited work on Negative Binomial regression that are rarely specified in full, but may be misleading if taken for granted. Our findings show disagreement with regards to scoring and information equations in the so-called dispersion parameter, or its reciprocal; and that there is no reliable expression for the corresponding element of the Fisher information matrix. 
	
	By elevating computational aspects that are typically overlooked, we make the following contributions: (i) reconcile notations that that adopt different parameterizations, or alternative approaches to the gamma function and its derivatives; (ii) address the absence of a reliable expression for the elements of the Fisher information matrix corresponding to parameters in the dispersion parameter; (iii) for practical relevance, implement a simplified numerical example underpinned by real-world data on clinical trials supply disruptions.
	
	A limitation of this work is its focus on cross-section Negative Binomial regression in its most basic form, although we touch on some extensions such as zero truncation and rate parametrization. Yet  the hoped for outcome is greater awareness that outsourcing proof to routinely-referenced sources is no guarantee, per se, that the underlying computational aspects can be safely overlooked or left unchallenged.

	%% Appendix 
	\begin{appendix}
		\section{Appendix}
		
		\setcounter{equation}{0}\renewcommand{\theequation}{A\arabic{equation}}
		\setcounter{table}{0}\renewcommand{\thetable}{A\arabic{table}}

		\subsection{Example applications}\label{secA1}	
		
		\begingroup	
		\renewcommand{\arraystretch}{0.6} 
		%\begin{longtable}[]{@{}llcccccc@{}}			
		%\begin{adjustbox}{max width=\linewidth}				
		\begin{threeparttable}				
			\caption{Example applications of Negative Binomial regression published in the \textit{Journal of Operations Management}, a flagship empirical management research journal, over two decades\tnote{1}.}\label{tab:tab1} 
			\begin{tabular}{lccccc}
				\toprule
				References & Count variable & \multicolumn{2}{l}{Equations\tnote{2}} & Software & \multicolumn{1}{c}{Methodology ref\tnote{3}} \\
				\cmidrule(l{3pt}r{3pt}){3-4}
				& & LL & PL & &   \\		
				\midrule
				\cite{JOM1_Boone} & Projects &   &   & & CT \\
				\cite{JOM2_Hahn} & Projects &   &   & & CT \\  
				\cite{JOM3_Park} & Gear shiftings &   & & STATA & HAU \\	
				\cite{JOM4_Rabinovich} & Sales &   &  & &  HAU \\	
				\cite{JOM5_Hofer} & Activities & &  & STATA & HAU \\
				\cite{JOM6_Bellamy14} & Patents & $\bullet $& & STATA &  CT, HIL \\
				\cite{JOM7_Bockstedt} & Submissions & &  & STATA &  \\
				\cite{JOM8_Steven} & Recalls &   & &  &  \\
				\cite{JOM9_Wagner} & Disruptions & & $\bullet$ & STATA &  CT, HIL \\
				\cite{JOM10_steven} & Recalls &   & &  & CT	\\
				\cite{JOM11_Ball} & Recalls &   &   & &  \\
				\cite{JOM11_restaurant} & Duration & &  & R (MASS) & HIL  \\
				\cite{JOM12_Lonati} & Citations &   &   & &  \\
				\cite{JOM13_cantor} & Orders & &  & SPSS  & \\
				\cite{JOM14_Chae} & Awards &   & &  &  \\		
				\cite{JOM16_tucker} & Injuries & &  &   & CT \\
				\cite{JOM17_park} & Sales & & $\bullet$ &   &  \\
				\cite{JOM18_Liu} & Participants & $\bullet $& & STATA & HIL \\
				\cite{JOM19_Scott} & Accidents &  &  & &  \\
				\cite{JOM20_singh} & Longevity & $\bullet $& &   & HIL  \\	
				\bottomrule						
			\end{tabular}		
			\begin{tablenotes}[para]
				\scriptsize	
				\item[1] Excludes applications as secondary method, for robustness checks only;			
				%	\item[2] A: \cite{JOM1_Boone}, B: \cite{JOM2_Hahn}, C: \cite{JOM3_Park}, D: \cite{JOM4_Rabinovich}, E: \cite{JOM5_Hofer}, F: \cite{JOM6_Bellamy14},  G: \cite{JOM7_Bockstedt}, H: \cite{JOM8_Steven}, I: \cite{JOM9_Wagner}; J: \cite{JOM10_steven}, 	K: \cite{JOM11_Ball}, L: \cite{JOM11_restaurant}, M: \cite{JOM12_Lonati}, N: \cite{JOM13_cantor}; O: \cite{JOM14_Chae}, P: \cite{JOM16_tucker}, Q: \cite{JOM17_park}, R: \cite{JOM18_Liu}, 	S: \cite{JOM19_Scott}; T: \cite{JOM20_singh};
				\item[2] LL: log-likelihood; PL: p.m.f. and mean function 
				\item[3] CT: \cite{Cameron_Trivedi_2013}, HAU: \cite{Hausman84}, HIL: \cite{Hilbe_2011}
			\end{tablenotes}
		\end{threeparttable}
		%\end{adjustbox}
		%\end{longtable}
		\endgroup
		\begingroup	
		\renewcommand{\arraystretch}{1}
		
		\endgroup
		
		\subsection{log-likelihood derivatives: regression coefficients}\label{sec:A2}	
		For completeness, we report the first and second derivative of Eq.~\eqref{eq:maxlik2}, and the associated element of the information matrix. We begin from the following gradient:
		
		\begin{align}
			\frac{\partial}{\partial\boldsymbol{\beta}}\ell\left(\boldsymbol{\beta}, \theta \right)  & = \sum_{i=1}^n\left\{y_i\mathbf{x}_i-\frac{\theta^{-1}\theta\mathbf{x}_ie^{\mathbf{x}_i^T\boldsymbol{\beta}}}{\left(1+\theta e^{\mathbf{x}_i^T\boldsymbol{\beta}}\right)}-\frac{\theta y_i\mathbf{x}_ie^{\mathbf{x}_i^T\boldsymbol{\beta}}}{\left(1+\theta e^{\mathbf{x}_i^T\boldsymbol{\beta}}\right)}\right\} \nonumber \\
			& = \sum_{i=1}^n\frac{y_i-e^{\mathbf{x}_i^T\boldsymbol{\beta}}}{\left(1+\theta e^{\mathbf{x}_i^T\boldsymbol{\beta}}\right)}\mathbf{x}_i \label{eq:nbgradbeta}
			%& = \nabla \ell\left(\boldsymbol{\beta}, \theta\right) 
		\end{align}
		
		Moving on to the Hessian, it may be computed as the first derivative of the gradient
		\begin{align}
			\frac{\partial^2}{\partial \boldsymbol{\beta}\partial \boldsymbol{\beta}^T}\ell{\left(\boldsymbol{\beta},\theta\right)} &=\sum_{i=1}^n{\frac{-\mathbf{x}_i e^{\mathbf{x}_i^T\boldsymbol{\beta}}-\theta\mathbf{x}_i e^{2\mathbf{x}_i^T\boldsymbol{\beta}}-y_i\theta\mathbf{x}_ie^{\mathbf{x}_i^T\boldsymbol{\beta}}+\theta\mathbf{x}_i e^{2\mathbf{x}_i^T\boldsymbol{\beta}}}{\left(1+\theta e^{\mathbf{x}_i^T\boldsymbol{\beta}}\right)^2}\mathbf{x}_i} \nonumber \\
			&=-\sum_{i=1}^n{\frac{\left(1+\theta y_i\right) e^{\mathbf{x}_i^T\boldsymbol{\beta}}}{\left(1+\theta e^{\mathbf{x}_i^T\boldsymbol{\beta}}\right)^2}\mathbf{x}_i\mathbf{x}_i^T} \label{eq:nbHbeta}
			%& = \mathbf{H}({\boldsymbol{\beta}}) 
		\end{align}
		
		where the outer product  $\mathbf{x}_i\mathbf{x}_i^T$ yields a matrix of dimensions $\left(p+1\right)\times\left(p+1\right)$, with $p$ the number of regressors. These results are analogous to those generally reported in the count regression literature \citetext{\citealp[e.g.,][Eq.~2.3 \& 2.5a]{Lawless87}; \citealp[Eq.~8.12 \& 8.15]{Hilbe_2011}}.
		
		The information matrix element corresponding to Eq.~\eqref{eq:nbHbeta} is minus its expectation. Unlike the Poisson specification, the presence of the random count-based variable in the derivative  is such that the ``expected'' information differs from the ``observed'' information, specifically: 
		
		\begin{equation}
			\textrm{E}\left[-\frac{\partial^2\ln\ell{\left(\boldsymbol{\beta},\theta\right)}}{\partial\boldsymbol{\beta}\partial\boldsymbol{\beta}^T}\right] =  \sum_{i=1}^n{\frac{e^{\mathbf{x}_i^T\boldsymbol{\beta}}}{1+\theta e^{\mathbf{x}_i^T\boldsymbol{\beta}}}\mathbf{x}_i\mathbf{x}_i^T} \label{eq:fibeta}
		\end{equation}
		
		What is more, differentiating w.r.t. $\theta$ the gradient in Eq.~\eqref{eq:nbgradbeta} above yields:
		
		\begin{equation}
			\textrm{E}\left[-\frac{\partial^2\ln\ell{\left(\boldsymbol{\beta},\theta\right)}}{\partial\boldsymbol{\beta}\partial\theta}\right] =
			\sum_{y_i=0}^{\infty}\left[\sum_{i=1}^n{\frac{y_i- e^{\mathbf{x}_i^T\boldsymbol{\beta}}}{\left(1+\theta  e^{\mathbf{x}_i^T\boldsymbol{\beta}} \right)^2}  e^{\mathbf{x}_i^T\boldsymbol{\beta}} \mathbf{x}_i}\right]\textrm{Pr}\left(Y_i\right)  = \mathbf{0} \label{eq:fibetaCross}
		\end{equation}
		
		Eqs.~\eqref{eq:fibeta}-\eqref{eq:fibetaCross} are in agreement with results given in widely-cited references \citetext{\citealp[e.g.,][Eq.~3.31 and 3.34]{Cameron_Trivedi_2013}; \citealp[Eq.~2.7a and 2.7b]{Lawless87}}.

		\subsection{Illustrative data}\label{sec:A3}
		
		\begingroup	
		\renewcommand{\arraystretch}{0.6}         
		\begin{threeparttable}	        
			\caption{Streamlined numerical example, clinical trial supply}\label{tab:tab2}         
			\begin{tabular*}{\linewidth}{@{\extracolsep{\fill}} lrrrrr}
				\toprule
				\multicolumn{1}{l}{Study ID} & \multicolumn{1}{l}{Target variable } & \multicolumn{4}{l}{Explanatory variables} \\
				\cmidrule(l{3pt}r{3pt}){3-6}
				& \multicolumn{1}{l}{DSR}  & \multicolumn{1}{l}{BIO} & \multicolumn{1}{l}{DUR} & \multicolumn{1}{l}{CLI} &  \multicolumn{1}{l}{SUB}\\
				\midrule
				PROT\_001 & 6 & 0 & 108.07 & 54 & 310\\
				PROT\_002 & 1 & 1 & 37.08 & 16 & 35\\
				PROT\_003 & 14 & 0 & 81.63 & 112 & 632\\
				PROT\_004 & 2 & 0 & 81.60 & 119 & 618\\
				PROT\_005 & 1 & 1 & 28.96 & 144 & 395\\
				\addlinespace
				PROT\_006 & 2 & 1 & 30.61 & 416 & 1255\\
				PROT\_007 & 10 & 0 & 69.63 & 120 & 1053\\
				PROT\_008 & 3 & 0 & 53.42 & 30 & 97\\
				PROT\_009 & 0 & 0 & 28.67 & 122 & 687\\
				PROT\_010 & 1 & 1 & 29.16 & 187 & 397\\
				\addlinespace
				PROT\_011 & 1 & 1 & 15.78 & 134 & 277\\
				\bottomrule
			\end{tabular*}
			\begin{tablenotes}[para]
				\scriptsize	
				DSR: count of disruptions; BIO: drug substance is biological rather than chemical;  DUR: trial duration; CLI: n. of clinics; SUB: subjects enrolled.		
			\end{tablenotes}
		\end{threeparttable}
		\endgroup
		\begingroup	
		\renewcommand{\arraystretch}{1}
		\endgroup

	\end{appendix}
	%% end of appendix

	%\textsc{\phantomsection\label{supplementary-material}
	%\bigskip
	%\begingroup		
	%	{\bf Online Supplementary Materials}: \texttt{R} scripts to implement the simplified example in Sec. \ref{sec-example}, are made available on \texttt{GitHub} at \url{https:\\} 
	%\endgroup}

	%\vfill
	%% if your bibliography is in bibtex format, uncomment commands:  
	\begingroup
	\setlength{\baselineskip}{15.5pt}
	\bibliography{NB_TAS_refs}       % Bibliography file (usually '*.bib')	
	\endgroup
	\end{document}